\documentclass[sn-mathphys,iicol]{sn-jnl}

\jyear{2021}%

\theoremstyle{thmstyleone}%
\theoremstyle{thmstyletwo}%

\theoremstyle{thmstylethree}%

\usepackage{amsthm}
\usepackage{widetext}

\begin{document}

\title[Article Title]{ Wavelength dependence of laser-induced excitation dynamics in silicon}


\author*[1]{\fnm{Prachi} \sur{Venkat}}\email{venkat.prachi@qst.go.jp}

\author[1,2]{\fnm{Tomohito} \sur{Otobe}}\email{otobe.tomohito@qst.go.jp}

\affil*[1]{\orgdiv{KPSI}, \orgname{National Institutes for Quantum Science and Technology}, \orgaddress{\street{8-1-7, Umemidai}, \city{Kizugawa}, \postcode{619-0125}, \state{Kyoto}, \country{Japan}}}
\affil[2]{\orgdiv{Photon Science Center}, \orgname{The University of Tokyo}, \orgaddress{\street{7-3-1, Hongo}, \city{Bunkyo-ku}, \postcode{111-8656}, \state{Tokyo}, \country{Japan}}}


\abstract{Effect of laser wavelength on the carrier-phonon dynamics and damage threshold of silicon is studied numerically. 
Laser excitation dynamics in silicon is studied using Three-Temperature Model (3TM). 
We consider the evolution of electron, hole, and lattice temperatures separately and including band-gap re-normalization effect on optical properties of silicon.
Finite Difference Time Domain method is used to model the laser field. 
Damage threshold calculated using the 3TM is in reasonable agreement with the experiments. Our results indicate that the competition of inter-band excitation, plasma heating, and electron-phonon relaxation process defines the damage threshold for various wavelengths and pulse durations.}




\maketitle

\section{Introduction}
\label{sec1}
Laser processing studies of semiconductors like silicon are crucial for the practical applications in the field of nano-fabrication \cite{Gattass2008,Stoian-2020}. Interaction of intense, femto-second pulses with a target can lead to high resolution in transfer of energy, with minimum damage to the surrounding area \cite{Sugioka2014}. Laser excitation of semiconductors involves complex physics of photo-absorption, impact ionization, re-combination and re-distribution of energy. It is the dynamics of the interaction that ultimately leads to structural changes through thermal and non-thermal effects.

Numerical modeling allows us to study the dynamics of laser excitation in silicon in detail. The onset and nature of damage can be better understood by taking into account all the processes occurring within the lattice and their respective time frames. 
Theoretical modeling of such interactions usually involves the use of Two-Temperature model (TTM), which is a widely implemented approach to study the evolution of electron and lattice dynamics during laser excitation. TTM was initially developed to study the dynamics in metals \cite{Anisimov1974} and then extended to study excitation in semiconductors \cite{Agassi-1984,VanDriel-1987}. More recently, a self-consistent density-dependent TTM (nTTM) has been implemented to study the evolution of carrier densities, temperatures and lattice temperature in the material \cite{Chen-2005}.

The dynamics following the laser excitation involves a progression of carrier excitation, relaxation and electron-hole-phonon dynamics.
The electron and hole relaxation in conduction and valence bands is observed to be faster than relaxation of the system as a whole \citep{Zurch2017}. Electron-hole scattering frequency decreases significantly with the increase in electron temperature \cite{Terashige-2015}. The calculations presented using first-principles numerical simulation show that electron and hole energies evolve differently, depending on the band structure. The energy difference between electrons and holes suggests that quasi-temperatures are different in the conduction band (CB) and valence band (VB) \citep{Otobe-2019}. Also, the decreased electron-hole scattering frequency indicates that it is important to take quasi-temperatures of electrons and holes into consideration. 
A modified nTTM treating the evolution of  quasi-temperatures in CB, VB and lattice (3TM) has been proposed in our previous work \cite{3TM-2022}.

\begin{table}[h]
    \centering
    \caption{List of symbols and notations used in the text}
    \hrule
    \vspace{0.05cm}
    \hrule
    \begin{tabular}{l | l }
    $n_{e(h)}$ & Carrier density  \\
    $n_0$ & Density of valence electrons \\
    $\eta_{e(h)}$ & Reduced quasi Fermi-level \\
    $q_{e(h)}$ & Carrier charge; $q_e=-q$; $q_h=q$ \\
    $\mu_{e(h)}$ & Electron(hole) mobility \\ 
    $T_{e(h)}$ & Electron(hole) temperature \\
    $T_l$ & Lattice temperature \\
    $k_B$ & Boltzmann Constant \\
    $D_{e(h)}$ & Diffusion coefficient \\
    $\vec{F}$   & Electric field due to e-h separation \\
    $\vec{j}_{e(h)}$ & Charge current \\
    $\vec{w}_{e(h)}$ & Energy current\\
    $\alpha$ & Single photon absorption coefficient \\
    $m_{r,e(h)}$ & $m^{*}_{e(h)}/(m^{*}_{e(h)}+m^{*}_{h(e)})$ \\
    $\alpha_f$ & Free carrier absorption coefficient \\
    $\beta$ & 2-photon absorption coefficient \\
    $\chi_r$ & Real part of susceptibility ($\chi=(\epsilon-1)/4\pi$) \\
    $\lambda$ & Laser wavelength \\
    $t_p$ & Pulse duration in FWHM\\
    
    \end{tabular}
    \hrule
    \vspace{0.05cm}
    \hrule
    \label{tab1}
\end{table}

The 3TM is modeled to consider three sub-systems, electrons, holes and lattice, and their temperatures are calculated separately. The carrier densities are also calculated distinctively for electrons and holes. We treat the dynamics of laser field by solving one-dimensional Maxwell's equations, using the Finite Difference Time Domain (FDTD) approach. The effect of band re-normalization on the optical properties of silicon is also included \cite{Sokol-2000}. We use Drude model to calculate the complex dielectric function, and thereby include the effect of free carrier response in photo-absorption \cite{Silaeva_2012}.

\begin{sidewaystable}
    \begin{center}
    \caption{List of parameters}
    \hrule
    \vspace{0.05cm}
    \hrule
 \begin{tabular}{ l | l | l  }    
    $\kappa_{l}$ & Lattice thermal conductivity \cite{VanDriel-1987} & $1585T_l^{-1.23}$ W/(cm K)  \\
    $C_l$ & Lattice heat capacity \cite{VanDriel-1987} & $1.978 + 3.54\times10^{-4}T_l-3.68T_l^{-2}$ J/cm$^3$ \\
    $m^{*}_{e}$ & Electron DoS effective mass \cite{Lipp-2014,BARBER-1967} & $0.36m_e$ \\
    $m^{*}_{h}$ & Hole DoS effective mass \cite{Lipp-2014,BARBER-1967} & $0.81m_e$ \\
    $\mu_e^0$ & Electron mobility \cite{Meyer-1980,Lipp-2014} & $8.5\times10^{-3}$ m$^2$/Vs \\
    $\mu_h^0$ & Hole mobility \cite{Meyer-1980,Lipp-2014} & $1.9\times10^{-3}$ m$^2$/Vs \\
    $\gamma_e$ & Auger recombination coefficient \cite{Silaeva_2012,Dziewior-1977} & $2.3\times10^{-31}$ cm$^6$/s \\
    $\gamma_h$ & Auger recombination coefficient \cite{Silaeva_2012,Dziewior-1977} & $7.8\times10^{-32}$ cm$^6$/s \\
    $\theta_{e(h)}$ & Impact ionization coefficient \cite{VanDriel-1987} & $3.6\times10^{10}\exp(-1.5E_g/k_BT_{e(h)}) s^{-1}$ \\
    $\delta E_g$ & Change in band gap & $1.5\times10^{-8}n_e^{1/3}$ eV \cite{Sokol-2000}\\
    $E_g$ & Band gap function &  $1.16 -7.02\times10^{-4}\frac{T_l^{2}}{T_l + 1108}  - \delta E_g$ eV\\
    $\tau$ & e-ph relaxation time \cite{Yoffa-1981, Agassi-1984,VanDriel-1987,sjodin-1998,Harb-2006} & $\tau_0(1 + (n_e/(8\times10^{20}))^2)$ \\
    $\tau_0$ & e-ph relaxation time constant \cite{sjodin-1998,Chen-2005} & 240 fs \\
  \end{tabular}
    \hrule
    \vspace{0.05cm}
    \hrule
    \label{tab2}
    \end{center}
\end{sidewaystable}

In order to understand and define damage in silicon, we can assume several processes that may be responsible for causing permanent structural changes. In our previous study, we have compared the calculated thresholds for possible damage mechanisms with the experimental damage thresholds to understand the nature of damage in silicon and the interaction conditions affecting it \cite{3TM-2022}.
Changing the wavelength of incident pulse can lead to significant changes in the dynamics of the interaction \cite{Gallais-2015}. Since the photon energy changes, the excitation mechanisms involving single- and two- photon absorption may be affected, in turn affecting the carrier density and temperatures. 
Apart from photon energy, the penetration depth for single- and two-photon absorption in silicon changes with the wavelength and can lead to different degrees of damage and structural changes. 

In this paper, we study the interaction of laser pulse of different wavelengths with a silicon film of fixed thickness and the effect on damage threshold. The paper is organized as follows: Section \ref{sec2} includes the details of the numerical model, sec. \ref{sec3} consists of the results and discussion and sec. \ref{sec4} summarizes the study.

\section{Numerical Scheme}
\label{sec2}
Laser irradiation of Silicon target leads to excitation of the electrons from valence band to conduction band and a complex interplay of carrier and lattice dynamics, ultimately leading to thermal equilibrium. Single- and two-photon absorption, impact ionization and Auger recombination occur during laser excitation of silicon and affect the temporal evolution of transient carrier densities. Carrier-phonon interaction is responsible for the re-distribution of energy within the carrier and lattice system.

The present model is similar to nTTM \cite{Chen-2005}, with some key differences. Firstly, the system is considered to consist of three sub-systems viz., electrons, holes and the lattice. The electron and hole densities are calculated separately, as are the electron, hole and lattice temperatures. Also, the effect of band structure re-normalization is taken into account while calculating the dielectric function and single- and two-photon absorption coefficients.The model is explained in detail in Ref. \cite{3TM-2022}. Tables \ref{tab1} and \ref{tab2} list the notation and values for various parameters and variables used.

The time-evolution of electron and hole densities, $n_e$ and $n_h$ is described as:
\begin{equation}
    \begin{split}
    \frac{\partial n_{e(h)}}{\partial t} & = \frac{\alpha I}{\hbar \omega_0} + \frac{\beta I^2}{2\hbar \omega_0} - \gamma_e n_en_en_h -\gamma_h n_hn_hn_e \\
     & + \frac{1}{2}(\theta_e n_e + \theta_h n_h) + \nabla D_{e(h)}\cdot\vec J_{e(h)} \\
     & + D_{e(h)}\nabla\cdot \vec J_{e(h)} \\ 
     & -(+) \mu_{e(h)}\nabla\cdot n_{e(h)}\vec{F} -(+) \mu_{e(h)}\nabla n_{e(h)}\cdot\vec{F}\\ & -(+) \mu_{e(h)} n_{e(h)} \nabla\cdot \vec{F}
    \end{split}\label{source}
\end{equation}
where $\omega_0$ is the laser frequency and $\alpha$ is the single-photon absorption coefficient for transition from VB to CB \citep{Green2008}. $\beta$ is the two-photon absorption coefficient for which we use the DFT calculation when $2\hbar\omega_0 > E_o$ \cite{Murayama-1995}, where $E_o$ is the optical gap.  Around the band gap energy we employ interpolation to the model described in Ref.\cite{Alan2007,Garcia_2006,Furey-2021}. $\gamma_{e(h)}$ is the Auger re-combination coefficient \citep{Silaeva_2012} and $\theta_{e(h)}$ is the impact ionization coefficient \citep{Chen-2005}. Equation \ref{source} also includes the effect of spatial charge distribution and the associated electric field, and $J_{e(h)}, D_{e(h)}$ and $\vec{F}$ are the charge current, diffusion coefficient and the electric field induced by the electron--hole separation, respectively \cite{3TM-2022}.

The total dielectric function along with the effect of band structure re-normalization \citep{Sokol-2000} is expressed by 
\begin{equation}
    \epsilon(\omega) = 1 + \frac{n_0 - n_e}{n_0}\epsilon_L(\omega + \delta E_g/\hbar) + \epsilon_D(\omega) 
\end{equation}
where $n_0$ is the density of valence electrons.  $n_e$ and $n_h$ are nearly the same due to the effect of $\vec{F}$ and can be approximated at $n_e$. $\epsilon_L(\omega)$ is the innate dielectric function, $\delta E_g$ represents the band re-normalization by carrier density, and $\epsilon_D$ is the complex  dielectric function calculated from Drude model \cite{3TM-2022}.
The total one-photon absorption coefficient including free-carrier absorption is calculated using the complex dielectric function \cite{3TM-2022}.
The temperature dependent optical parameters of silicon are referred to from Ref.\citep{Green2008}.

The temperature evolution of the electron, hole and lattice sub-systems is calculated as:
\begin{equation}
\begin{split}
    & C_{e(h)}\frac{\partial T_{e(h)}}{\partial t} =  m_{r,{e(h)}}(\alpha_f I + \beta I^2) \\ 
    & + E_g\gamma_{e(h)} n_{e(h)} n_{e(h)} n_{h(e)} \\
    & -\frac{C_{e(h)}}{\tau}(T_{e(h)}-T_l)
-\nabla \cdot \vec{w}_{e(h)} \\
& -\frac{\partial n_{e(h)}}{\partial t} \left(m_{r,{e(h)}}E_g 
+ \frac{3}{2}k_B T_{e(h)} H_{-1/2}^{1/2}(\eta_{e(h)}) \right) \\ & - m_{r,{e(h)}}n_{e(h)}\left( \frac{\partial E_g}{\partial T_l}\frac{\partial T_l}{\partial t}
+  \frac{\partial E_g}{\partial n_{e(h)}}\frac{\partial n_{e(h)}}{\partial t}\right),
\end{split}
\label{Te}
\end{equation}

\begin{equation}
    C_l\frac{\partial T_l}{\partial t}=-\nabla\cdot(\kappa_l \nabla T_l)+\frac{C_e}{\tau}(T_e-T_l)+\frac{C_h}{\tau}(T_h-T_l).
    \label{Tl}
\end{equation}
The third and fourth terms in Eq.~(\ref{Te}) account for the loss of energy due to electron-lattice interaction and energy current. The last two terms on right hand side include the changes in carrier density and band gap energy. Here, $H_{\xi}^{\zeta}(\eta)=F_{\zeta}(\eta)/F_{\xi}(\eta)$ and  $F_{\xi}(\eta)$ is the Fermi integral. The heat capacities $C_{e(h)}$ are calculated from the carrier densities and temperatures. 
$T_l$ is calculated following the empirical model, where the term for carrier temperature is replaced by terms for electron and hole temperatures, as described in Eq. (\ref{Tl}). Here, $\kappa_l$ is the thermal conductivity \cite{3TM-2022}.
 
The propagation of the laser pulse is described by solving the Maxwell's equations using FDTD method. Mur's absorbing boundary condition is employed to prevent reflection from the boundary \citep{Mur-1981}. Also, the electric field is considered to be complex for the calculation of laser intensity, so as to ensure a non-zero field at all points in time and space. Assuming a one-dimensional system, the electric field is:
\begin{equation}
    E(x,t) = E_0(x,t)\exp[i\omega t]
\end{equation}
Here $E_0$ includes Gaussian envelope defining the shape of the pulse as $\exp[-(t-4T)^2/T^2]$, where $T=t_p/(4\sqrt{\ln 2})$, $t_p$ being the FWHM pulse duration.
\begin{equation}
    I(x,t) = \frac{c}{8\pi}\Re[\sqrt{\epsilon}]\lvert E_{0}(x,t) \rvert^2
\end{equation}
where $I(x,t)$ is the laser intensity. 
Evaluation of charge current induced by the laser field is a crucial part of the module. We calculate the current with and without excitation i.e., for photo-absorption and dielectric response. For dielectric response, $j_0(x,t)$ is calculated as:
\begin{equation}
    j_0(x,t) = \chi_r(\omega)\frac{\partial P(x,t)}{\partial t} = -\chi_r(\omega)\frac{\partial^2A(x,t)}{c\partial t^2}
\end{equation}
where $A(x,t)$ is the vector potential,  $P$ is the polarization and $\chi_r$ is the real part of susceptibility ($\chi = (\epsilon-1)/4\pi$).
The Maxwell's equation thus, becomes 
\begin{equation}
   \frac{1}{c^2}(1+4\pi\chi_r(\omega))\frac{\partial^2A(x,t)}{\partial t^2} - \frac{\partial^2A(x,t)}{\partial x^2} = \frac{4\pi}{c}j(x,t)
\end{equation}
where
\begin{equation}
    j(x,t) = (\alpha_f(\omega) + \beta(\omega)I(x,t))\frac{c\Re[\sqrt{\epsilon}]}{4\pi}E(x,t)
\end{equation}
is the current associated with photo-absorption.

\begin{figure}[t]
    \centering
    \includegraphics[width=0.5\textwidth]{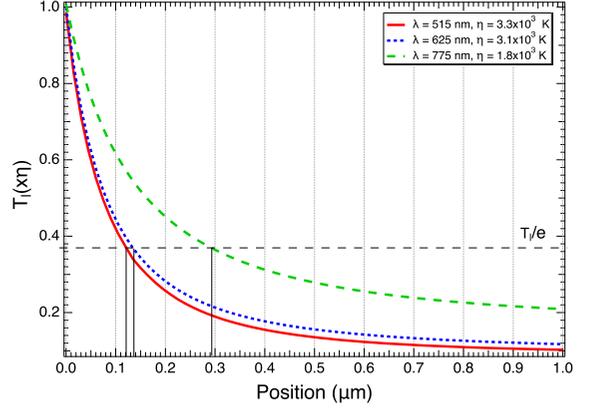}
    \caption{Variation of lattice temperature with the depth of target is plotted at the last time step for different $\lambda$.}
    \label{fig1}
\end{figure}


\section{Results and discussion}
\label{sec3}
We study the interaction of laser pulse of wavelengths ($\lambda$) 515 nm ($\hbar\omega_0=2.41$ eV), 625 nm ($\hbar\omega_0=1.98$ eV) and 775 nm ($\hbar\omega_0=1.6$ eV) with silicon film of thickness 60 $\mu$m. Since the penetration depth differs depending on the wavelength, we choose a film thickness where the reflection from the rear surface does not cause interference with the EM-field within the film. The corresponding penetration depths for single-photon absorption ($1/\alpha$) in case of $\lambda~=$ 515 nm, 625 nm and 775 nm are $\sim1 \mu$m, $\sim5 \mu$m and $\sim10 \mu$m respectively \cite{Green-1995}.


Figure \ref{fig1} shows the lattice temperature gradient within the silicon film at the last time step of the simulation for different $\lambda$. The pulse duration and incident fluence are 350 fs and 0.18 J/cm$^2$.  The temperature is normalized by the factor of $\eta$. The maximum lattice temperature on the surface of the film is given by $\eta$, which decreases with the decrease in photon energy. The penetration depth for single-photon absorption is different in case of each $\lambda$. We also look at the evolution of lattice temperature with position and the depth at which $T_l$ falls to $1/e$ of the maximum value. This processed depth for $T_l/e$ is 0.12 $\mu$m in case of $\lambda~=$ 515 nm, 0.14 $\mu$m for $\lambda~=$ 625 nm and 0.3 $\mu$m for $\lambda~=$ 775 nm. The change in processed depth indicates the effect of wavelength on the dynamics, which may be due to the different photon energies, affecting the absorption mechanism and plasma response.
The two-photon absorption coefficient is 37.6 cm/GW \cite{Murayama-1995}, 36.9 cm/GW \cite{Murayama-1995} and 1.9 cm/GW \cite{Alan2007} for $\lambda~=$ 515 nm, 625 nm and 775 nm respectively. Both single- and two-photon absorption processes become less intense with the increasing wavelength.
$1/\beta$ for the three wavelengths is 0.027 GW/cm, 0.027 GW/cm and 0.526 GW/cm, respectively. The processed depth increases with increasing $\lambda$, which seems inconsistent with the decrease in $\beta$ and $\alpha$. This indicates that while photon absorption becomes less intense with increase in $\lambda$ from 515 nm to 775 nm, plasma heating may observe an increase. The absorption of energy by the plasma can be considered responsible for the increased processed depth in case of $\lambda~=$ 775 nm.


The variation in the dynamics and processed depth with $\lambda$ would also manifest as change in the damage threshold, and has been observed in experimental studies as well \cite{Sokol-2000,Pronko-1998,Bonse-2002,Smirnov2018,Gallais-2015}. The effect of $\lambda$ variation can also potentially help control the processing depth in the target. 
The damage or permanent structural alterations in the sample can be studied in different ways \cite{Allen-2003,Bonse-2002,Izawa_2006,Mirza2016,Thorstensen-2012}.
The possible processes for damage and the calculation of their threshold has been discussed in detail in our previous work \cite{3TM-2022}.
We now consider three possible processes during the laser excitation of silicon, that may be responsible for causing damage.

\begin{figure}[h!]
    \centering
    \includegraphics[width=0.5\textwidth]{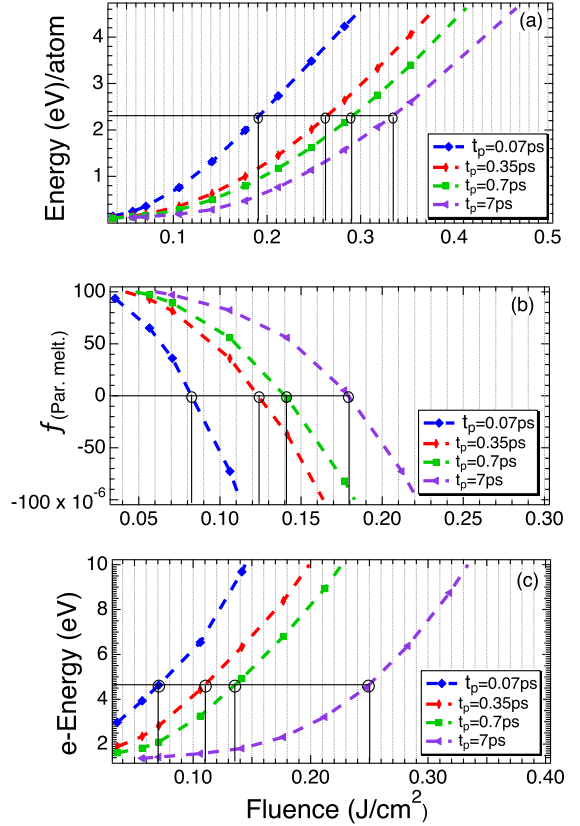}
    \caption{Plots of (a) Maximum energy per atom (b) $f_{(Par.~melt.)}$ and (c) electron (e) - energy vs. the incident laser fluence for different $t_p$. The circled point on each plot depicts the threshold for that pulse duration. In this case, $\lambda=625$ nm.}
    \label{fig2}
\end{figure}

Thermal melting can be considered significant in causing damage in the silicon lattice. The  threshold ($F_{TM}$) can be calculated for the onset of melting as lattice temperature reaches melting point of 1687 K (partial melting), or when bond configuration changes to tetrahedral after absorption of the total latent heat from the carriers (complete melting) \cite{Korfiatis-2007}. 
The two different thresholds for melting lead to different understanding of the onset of damage. 
The bond-breaking energy can also be considered as a reasonable candidate for the damage threshold \cite{Sato-2014}. 
 Threshold for breaking of bonds ($F_{BB}$) is  calculated as the fluence at which energy per atom reaches 2.3~eV i.e., half the cohesive energy of a silicon crystal \cite{Lutrus-1993}. 
Emission of electrons from the surface into vacuum is also crucial in studying damage mechanisms. When the electron temperature is high, electron emission (e-emission) from the surface into the vacuum causes impulsive Coulomb force within the lattice. This force tends to de-stabilize the lattice structure, and in extreme cases causes Coulomb explosion \cite{Roeterdink-2003}. 
The photon energy in the three cases being studied in this work is much lower than the work function of silicon (4.65~eV), therefore photo-emission by single- and two-photon absorption is not possible.
The Fermi-Dirac distribution suggests e-emission is caused as a result of thermal effects. We define the e-emission threshold ($F_{EE}$) as $E_g+1.5k_B T_e= 4.65$~eV. 
It may be noted that this assumption considers e-emission as the mere trigger for Coulomb forces that may be responsible for the actual damage.
A comparison of the calculated thresholds with experimental data from Allenspacher \textit{et.al} in our previous study showed that e-emission threshold coincided quite well with the damage threshold for longer $t_p$ \cite{3TM-2022}.

\begin{figure}[h!]
\includegraphics[width=0.5\textwidth]{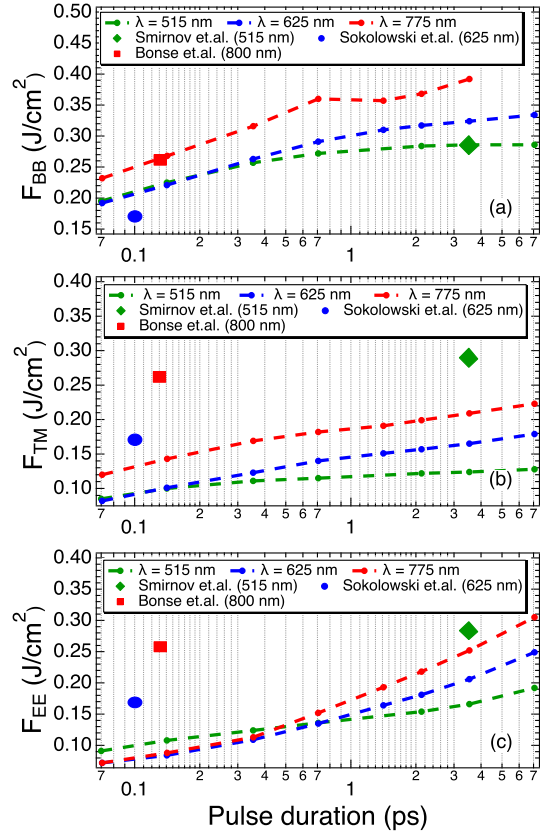}
    \caption{Calculated threshold for (a) breaking of bonds, (b) partial melting (c) and e-emission for different $\lambda$ and $t_p$ is plotted, along with the experimental thresholds.}
    \label{fig3}
\end{figure}

Figure \ref{fig2} shows the plots for (a) energy per atom, (b) function to calculate the onset of melting ($f_{(Par.~melt.)}$) \cite{Korfiatis-2007} and (c) electron energy for the case of $\lambda~=$ 625 nm and different $t_p$. The threshold for each process is determined from the plot as the fluence at which the plot reaches the threshold condition. $F_{BB}$ is marked on each $t_p$ plot where the maximum energy per atom is 2.3 eV. In case of partial melting, $F_{TM}$ is marked where the function \textit{f}$_{(Par.~melt.)}$ becomes zero. In case of e-emission, the fluence at which the average energy of electrons is 4.65 eV, is marked as $F_{EE}$. 
The thresholds for the three processes obtained in this manner are then plotted with pulse duration ($t_p$) in fig. \ref{fig3}.

Figure \ref{fig3} shows the calculated $F_{BB}$, $F_{TM}$ and $F_{EE}$ for pulses of three different wavelengths. The calculated thresholds are plotted for different $t_p$. Since the lattice temperature evolution is similar for 515 nm and 625 nm as shown in fig. \ref{fig1}, $F_{BB}$ and $F_{TM}$ are nearly the same, as long as $t_p < 0.4$ ps. 
As $t_p$ increases well beyond the e-ph interaction time and the relaxation time for electron-phonon dynamics,the dynamics and energy transition get affected, and the thresholds are not the same for the three cases of $\lambda$. 

In fig. \ref{fig3} (c), $F_{EE}$ for different wavelengths show a similar trend. For 775 nm pulse, the e-emission threshold is comparable with the thresholds for shorter $\lambda$, for $t_p < 0.4-0.5$ ps. For longer $t_p$, the thresholds become markedly different for the three cases. The dependence of threshold on $t_p$ is increased \cite{3TM-2022} and combined with the difference in photon energy, the e-emission threshold for longer pulses is different for varying $\lambda$. It can also be observed that $F_{EE}$ is lower for the lower photon energies of 1.6 and 1.98 eV. 
This may be a result of the change in plasma frequency with increasing wavelength. The resonant absorption depends on the plasma frequency and critical density for a particular case of $\lambda$ and this may be responsible for the lower e-emission threshold for longer wavelength.

On comparing the calculated values of $F_{BB}$, $F_{TM}$ and $F_{EE}$, it was seen that $F_{BB}$ coincided most with the experimental data. Table \ref{tab3} shows the experimental threshold for different cases of $\lambda$ and the calculated threshold for breaking of bonds. The experimental data is also indicated in fig. \ref{fig3}. 
The damage threshold is defined differently in the experimental studies, making the comparison of calculated and experimental data for different studies challenging. 
\begin{table}[h]
    \centering
    \caption{Comparison of calculated threshold  for breaking of bonds (F$_{BB}$ J/cm$^2$) with the experimental data (F$_{exp.}$  J/cm$^2$)}
    \hrule
    \vspace{0.05cm}
    \hrule
    \begin{tabular}{c | c | c |c | c}
       $\lambda$ (nm) & $t_p$ (ps) & F$_{exp.}$ & F$_{BB}$ & $\Delta (\%)$ \\
       \hline
       515 \cite{Smirnov2018} & 3.5 & 0.286 & 0.286 & 0.0 \\
       \hline
       625 \cite{Sokol-2000} & 0.1 & 0.17  & 0.21 & 23.5 \\
       \hline
       775 \cite{Bonse-2002} & 0.13 & 0.26 & 0.264 & 1.5 \\
    \end{tabular}
    \hrule
    \vspace{0.05cm}
    \hrule
    \label{tab3}
\end{table}
In case of $\lambda~=$ 625nm, the experiment observed the melting threshold. The calculated threshold for melting as well as breaking of bonds showed $\sim 23\%$ error \cite{Sokol-1995}.
For $\lambda~=$ 775nm, the experimental threshold was calculated by plotting the diameter of the damaged spot with the incident fluence and extrapolating the plot to obtain the fluence when diameter became zero \cite{Bonse-2002}. 
The agreement of experimental data with breaking of bonds may be due to the definition of threshold, which is defined when the damaged spot appears.
The experiments where damaged area is considered means that the damage probability is 100 $\%$. On the other hand, in the experimental study by Allenspacher et.al., the damage threshold is determined when the damage probability is 0 $\%$ \cite{Allen-2003}. Due to this difference in definitions, the experimental data in former case agrees with calculated $F_{BB}$, and in latter case it agrees with calculated $F_{TM}$ and $F_{EE}$ \cite{3TM-2022}.

In case of $\lambda~=$ 515 nm, the experimental data reported by Smirnov \textit{et. al} consists of damage threshold over a range of pulse duration \cite{Smirnov2018}. Smirnov \textit{et. al} defined the threshold from the radius of the ablation crater by considering it as a function of the natural logarithm of the energy ($R^2(lnE)$). Ablation threshold is $F_{th} = E_{abl}/\pi W_{abl}^2$ where $E_{abl}$ is ablation threshold and $W_{abl}$ is the focal spot at 1/e of the energy density. Comparison of our calculated threshold with Smirnov \textit{et.al} data showed that the calculations were in agreement with experiment for pulse duration longer than 1 ps, and the data for a particular case of $t_p=3.5$ ps is shown in table \ref{tab3}.
For $t_p < 1$ ps, the calculated threshold did not match the experimental threshold. For shorter pulses of $\sim0.6$ ps, the experimental threshold is $\sim0.4$ J/cm$^2$, which is much higher than the calculated threshold.  Smirnov \textit{et.al} explain the high threshold for shorter $t_p$ as a result of faster ambipolar diffusion, as opposed to shorter diffusion stage and slower energy transition for longer $t_p$. 
One possibility for the disagreement  between our results and Smirnov \textit{et. al} data for $t_p<1$~ps is the different definition of threshold. 
The morphology of the ablation spots and the conditions for their formation need to be studied in detail to calculate more accurate thresholds. For example, Moser \textit{et. al} have reported that the hydrodynamic processes affect ablating spot size \cite{Moser18}. 

\section{Summary}
\label{sec4}

We studied the interaction of laser pulses of varying wavelength with silicon film. The dynamics show a significant change with the wavelength, which in turn affects the damage threshold. While the bond breaking and thermal melting threshold for the case of 775 nm pulse are significantly higher due to low photon energy, the e-emission threshold for the three wavelengths is comparable due to the plasma response. The  $t_p$ dependence of e-emission threshold increases when $t_p$ exceeds 0.4 ps, which may be due to the effect of carrier-phonon dynamics.

Our calculations indicate that the wavelength dependence is determined by the competition between inter-band transitions, plasma heating, and electron-lattice interactions. The first two are expected to have an effect for relatively short pulse duration, while the latter becomes more pronounced at longer pulse duration.
The calculated threshold for breaking of bonds is observed to coincide the most with the experimental data.

The agreement of calculations with experimental data maybe also be due to the treatment of photo-absorption in our model. We calculate the single- and two-photon absorption coefficients as a spectrum, so as to include the effect of changing laser frequency. The study for longer wavelengths can also be done using this model, although the effect of three-photon absorption must be taken into account. 

\section*{Acknowledgments}
This research is supported by MEXT Quantum Leap Flagship Program (MEXT Q-LEAP) under Grant No.  JPMXS0118067246. 
This research is also partially supported by JST-CREST under Grant No. JP-MJCR16N5.
The numerical calculations are carried out using the computer facilities of the SGI8600 at Japan Atomic Energy Agency (JAEA).



\bibliography{Refs}

\end{document}